**AI-Assisted Pleural Effusion Volume Estimation from Contrast-Enhanced CT Images**

Sanhita Basu[1], Tomas Fröding[2], Ali Teymur Kahraman[1], Dimitris Toumpanakis[3], Tobias Sjöblom[1]

[1]Department of Immunology, Genetics and Pathology, Uppsala University, Uppsala, Sweden

[2]Department of Radiology, Nyköping Hospital, Nyköping, Sweden

[3]Karolinska University Hospital, Stockholm, Sweden

**Abstract**

**Background:** Pleural Effusions (PE) is a common finding in many different clinical conditions, but accurately measuring their volume from CT scans is challenging.

**Purpose:** To improve PE segmentation and quantification for enhanced clinical management, we have developed and trained a semi-supervised deep learning framework on contrast-enhanced CT volumes.

**Materials and Methods:** This retrospective study collected CT Pulmonary Angiogram (CTPA) data from internal and external datasets. A subset of 100 cases was manually annotated for model training, while the remaining cases were used for testing and validation. A novel semi-supervised deep learning framework, Teacher-Teaching Assistant-Student (TTAS), was developed and used to enable efficient training in non-segmented examinations. Segmentation performance was compared to that of state-of-the-art models.

**Results:** 100 patients (mean age, 72 years ± 28 [standard deviation]; 55 men) were included in the study. The TTAS model demonstrated superior segmentation performance compared to state-of-the-art models, achieving a mean Dice score of 0.82 (95% CI, 0.79 – 0.84) versus 0.73 for nnU-Net ($p < 0.0001$, Student's T test). Additionally, TTAS exhibited a four-fold lower mean Absolute Volume Difference (AbVD) of 6.49 mL (95% CI, 4.80 – 8.20) compared to nnU-Net's AbVD of 23.16 mL ($p < 0.0001$).

**Conclusion:** The developed TTAS framework offered superior PE segmentation, aiding accurate volume determination from CT scans.

**Introduction**

Pleural effusion (PE) can result from > 50 different pathological conditions, such as congestive heart failure, pneumonia, lung cancer, inflammatory diseases, acute pancreatitis, and trauma. The presence (1) and amount (2,3) of effusions are important indicators of disease progression. A small amount of fluid, 2 to 4 mL, is present in each pleural cavity under normal conditions (4). Computed tomography (CT) is the state-of-the-art method for identifying and evaluating PE. CT scans allow for measurement of the size of the effusion, but determining the exact amount of fluid present is challenging and time-consuming for the radiologist.

Initial studies used manual methods to measure the height, depth, and width of effusions from sagittal, transverse, and coronal planes (4,5). However, this approach is time-consuming and prone to inter-observer variability (6). As a result, effusions are most often only described qualitatively in radiology reports. Quantification using traditional image processing or atlas-based segmentation has had limited performance (7). Recently, an AI-based method for PE segmentation in CT volumes was developed based on nnU-Net (6,8,9). The primary impediment to research is the lack of a large and diverse annotated dataset that covers the wide range of appearances of pathological pleural fluid (10), which also makes it difficult to evaluate methods to determine which is the most effective for different types of PE.

Here, we devise and implement a new framework for semi-supervised learning, which we apply to PE detection and quantification in CT examinations and benchmark to the performance of state-of-the-art algorithms.

**Materials and methods**

In Semi-Supervised Semantic Segmentation, a model is trained on a combination of labelled and unlabelled data. The labelled data, denoted by $\mathcal{D} = \{x_i, y_i\}_{i=1}^{N}$, is limited in quantity, whereas the unlabelled data, denoted by $\mathcal{U} = \{x_i\}_{i=1}^{N}$, is much larger than the labelled data. The semantic segmentation network consists of a feature extractor ($f$) and a mask predictor ($g$) network. The main challenge of this process is to make the best use of a large amount of unlabelled data available. One commonly used approach to tackle this challenge is pseudo-labelling, as described in references (19,20) which involves assigning pseudo-labels to the unlabelled data based on model predictions in real-time. The Teacher-Student framework is a popular approach currently used in Semi-Supervised Segmentation (21,22)as shown in (Fig.2). It involves two models *viz.* a "teacher" and a "student" model. The teacher model is responsible for producing pseudo labels (represented by $\hat{t}$), while the student model learns from both the ground-truth labels ($y$) and pseudo labels ($\hat{t}$). The teacher model uses the Exponential Moving Average (EMA) of the student model to generate updated pseudo labels, allowing the student model to learn more comprehensive features, resulting in improved performance.

Despite its effectiveness, the pseudo-labelling approach can generate unreliable pseudo-labels, resulting in inaccurate mask predictions. To address the challenge, here we propose a new pseudo-labelling approach that we call the "Teacher-TA-Student" framework, which can effectively learn representative features from unlabelled data while also mitigating the negative effects of unreliable pseudo-labels. The proposed Teacher-TA-Student framework for semi-supervised semantic segmentation is shown in (Fig. 2). The learning procedure of the Teacher-TA-Student consists of the following three steps.

**Step 1:** The Teacher model ($t$) is used to create the voxel-wise pseudo segmentation mask ($\hat{y}^t$) for the unlabelled CT images from the unlabelled dataset $\mathcal{U}$. Since the Teaching Assistant ($ta$) will learn from the pseudo-segmentation masks generated by the Teacher, it is important to filter out or give lesser importance to those voxels where the Teacher is unsure. We have selected a confidence threshold ($\tau$) to filter out pseudo labels with low confidence, keeping only those with high confidence. Suppose for $i^{th}$ voxel of an unlabelled CT volume from $\mathcal{U}$, the teacher model prediction probability is $p_i^t$, then we often keep those voxels whose confidence value is greater than one threshold, and generate pseudo labels as,

$$\hat{y}_i^t = \begin{cases} p_i^t & \text{if } p_i^t > \tau_e; \\ \text{ignore} & \text{otherwise.} \end{cases}$$

Here $\tau_e$ is the confidence threshold at the $e^{th}$ epoch, which can be a constant or can be varied among epochs. This re-weighting mechanism highlights the pixels with high confidence and suppresses those with low confidence to further enhance the reliability of the selected labels. This strategy helps to reduce the negative impact of unreliable pseudo-labels on the training process. The unlabelled data that are assigned pseudo labels will be taken as auxiliary training data for $ta$, while the other unlabelled data will be ignored.

**Step 2:** The Teaching Assistant ($ta$) plays a vital role in our framework. Other approaches attempt to have the student model learn from both labelled and unlabelled data at the same time. However, we believe that treating ground-truth labels and pseudo-labels as equal could be dangerous, as unreliable pseudo-labels may mislead the mask prediction. To overcome this issue, we aim to separate the effects of pseudo-labels on the student model's feature extractor and mask predictor. Our $ta$ learns from the unlabelled data and transfers only the useful feature representation knowledge to the student model, protecting it from any negative impact caused by unreliable pseudo-labels. After being optimized on unlabelled data using pseudo-labels, the $ta$ model conveys the learned feature representation knowledge to the student model via Exponential Moving Average (EMA) as,

$$w^{ta} = w^{ta} - \alpha \frac{\partial \mathcal{L}^{ta}}{\partial w^{ta}}, \quad \text{and} \quad w_f^s(t) = \gamma w_f^s(t-1) + (1-\gamma) w_f^{ta}(t),$$

where $w^{ta}(t)$ and $w^s(t)$ represent the parameters of the Teaching Assistant and the student models at the $t^{th}$ iteration. The symbol $w_f$ represents the parameters of the feature extractor. $\alpha$ denotes the learning rate and $\gamma \in [0, 1]$ is a hyper-parameter in EMA. $\mathcal{L}^{ta}$ is the loss function of $ta$ (eqn.6) and measures the distance between two distributions using $KL$ divergence loss. Here the distributions are the class probabilities of the Teacher ($\hat{y}^t$) and the teaching assistant ($\hat{y}^{ta}$). By minimizing the $KL$ loss, we are teaching the $ta$ to make similar predictions to the $t$.

$$\mathcal{L}^{ta} = KL(\hat{y}^t || \hat{y}^{ta}) = \sum_i \hat{y}_i^t \log \hat{y}_i^t - \hat{y}_i^t \log \hat{y}_i^{ta}.$$

**Step 3:** Finally, training of the Student model ($s$) with ground truth labels is performed along with the EMA-based knowledge sharing between the Student and Teacher models. With the Teaching Assistant module, the student model in our framework is only required to learn from the ground truth / labelled data. The optimization of the student model can be formulated as,

$$w^s = w^s - \alpha \frac{\partial \mathcal{L}^s}{\partial w^s}, \quad \text{and} \quad w_f^t(t) = \gamma w_f^t(t-1) + (1-\gamma) w_f^s(t), \quad w_g^t(t) = \gamma w_g^t(t-1) + (1-\gamma) w_g^s(t)$$

Where $w^s(t)$ and $w^s(t)$ represent the parameters of the Student and the Teacher models at the $t^{th}$ iteration. The symbol $w_f$ and $w_g$ represent the parameters of the feature extractor and the predictor. $\alpha$ denotes the learning rate and $\gamma \in [0, 1]$ is a hyper-parameter in EMA. $\mathcal{L}^s$ is the loss function of student and measures the overlap between the segmentation prediction of the Student ($\hat{y}^s$) and ground truth segmentation ($y$) as,

$$\mathcal{L}^s = DICE(y, \hat{y}^s) = 1 - \frac{2 \sum_i y_i \hat{y}_i^s}{\sum_i y_i^2 + \sum_i \hat{y}_i^{s2}}$$

Finally, the teacher model will be selected as the ultimate model for making predictions during inference because it incorporates information from both labelled and unlabeled data gathered by the student model.

**Experimental Setup and Results**
**Data collection and annotation**

We collected retrospective CT Pulmonary Angiogram (CTPA) data from internal CTPA (11) and external PleThora (12) datasets (Fig. 3). The CTPA dataset comprises 700 retrospective CTPA scans obtained from 652 patients (with a median age of 72 years and a range from 16 to 100 years) at Nyköping Hospital in Sweden between 2014 and 2018. Among these scans, 281 examinations had PE, while the remaining 419 were normal or had other lung and heart-related pathologies. The collection and analysis of CTPA examinations were approved by the Swedish Ethical Review Authority (EPN

Uppsala Dnr 2015/023 and 2015/023/1). All CTPA data were anonymized and exported in Digital Imaging and Communications in Medicine (DICOM) format. Reference volumetric segmentation for PEs was generated using a two-step semi-automatic manual segmentation pipeline. Initially, a junior annotator with over three years of experience in medical image processing and analysis employed the semi-automatic segmentation tools available in the MITK Workbench to outline the PE regions. Subsequently, a senior radiologist reviewed and corrected each case. We selected a representative sample of 100 cases from the positive cohort for manual annotation.

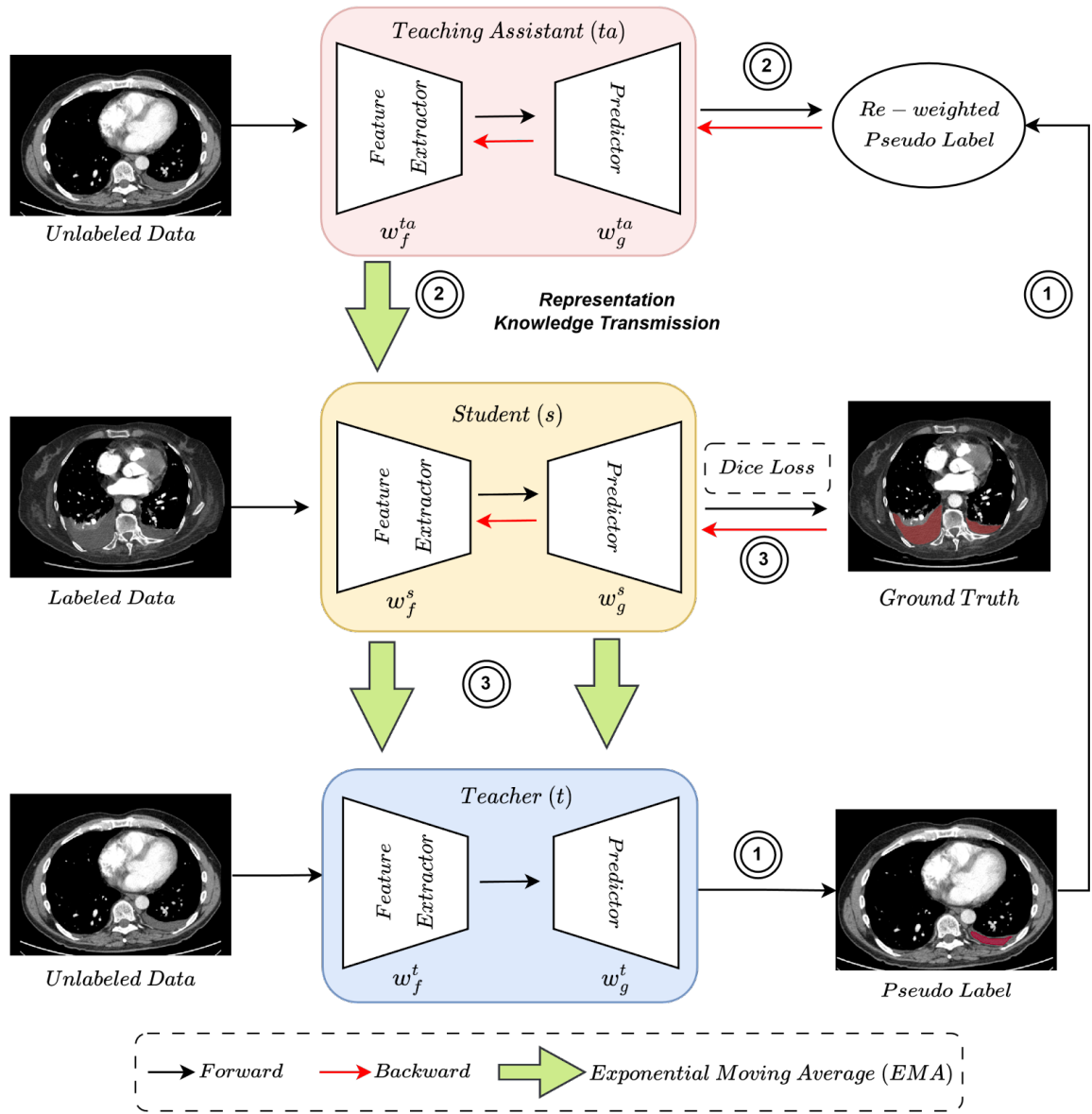

**Figure 1.** The step-by-step learning procedure of the Teacher-TA-Student model.

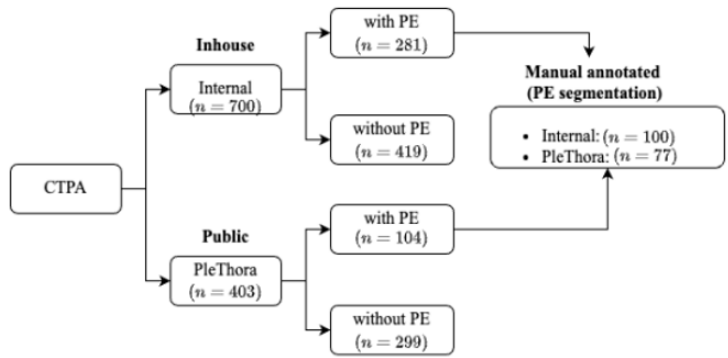

**Figure 2.** Internal and external datasets with the number of samples used for the experiments.

The selection criteria were based on the radiologist's recommendation, aiming to diversify the cases according to the volume, position, and complexity of the PE. The same two-step semi-automatic-manual segmentation pipeline was applied to generate manual annotations for these 100 cases. These 100 cases were used for model training along with samples from the negative cohort to improve the generalization of the AI models. PleThora[12] is a publicly available dataset (https://doi.org/10.7937/tcia.2020.6c7y-gq39) consisting of 402 CT scans from The Cancer Imaging Archive, where manual PE segmentations are available for 77 cases which were used for segmentation model validation and testing.

**Results**

We compared the proposed Teacher-TA-Student (TTAS) model with three other state-of-the-art models. The first was the conventional Teacher-Student model (TS) (14), utilized to evaluate the performance enhancement attributed to the proposed TA component. The second was the state-of-the-art nnU-Net model, which has achieved the best performance in segmenting Pleural Effusion (PE) from CT volumes (6). Finally, we also compared TTAS with the 3DU-Net (15) which is considered one of the best-performing methods for medical image segmentation. For model training, we used the 100 annotated sample cases from the in-house AIDA-CTPA dataset. Additionally, the remaining 181 sample cases were utilized for training the semi-supervised learning-based models such as the proposed Teacher-TA-Student (TTAS) and the Teacher-Student model (TS). For a comprehensive and transparent comparison, we adopted the approach outlined in (6), and assessed segmentation performance using the publicly available PleThora dataset (12). The CNN models were developed and trained using PyTorch 2.0.1 in Python 3.9. The experiments took place on a GPU server equipped with a sole NVIDIA Tesla V100, featuring 32 GB of memory.

The proposed TTAS model achieved a higher level of segmentation performance than the state-of-the-art methods (Table 1). The TTAS achieved a mean Dice score of 0.82, which is a considerable improvement from its competitor nnU-Net, which achieved 0.72. The mean Absolute Volume Difference (AbVD) provides a key measure to assess the quality of segmentation. For the quantification of PE volume, the TTAS model achieved about four-fold lower average error and SD than nnU-Net. The difference between TTAS and the competing methods with respect to these different metrics was statistically significant ($p<0.0001$, paired Student's t-test).

**Table 1. The Teacher-TA-Student model outperforms state-of-the-art models in key performance metrics.** Quantitative performance of different models on the PleThora dataset ($n = 77$). The best results are indicated in bold font. Up arrows, a higher value is preferred. Down arrows, lower value is preferred.

| Segmentation method | Performance metrics | | | | |
|---|---|---|---|---|---|
| | Dice (↑) | Precision (↑) | ASD (↓) | AbVD in mL (↓) | $p$-value |
| Teacher-TA-Student (TTAS) | 0.82±0.01 | 0.85±0.08 | 1.76±2.78 | 6.49±7.51 | – |
| Teacher-Student model (TS) | 0.78±0.12 | 0.79±0.12 | 2.78±3.27 | 11.25±21.83 | < 0.0001 |
| nnU-Net | 0.73±0.12 | 0.72±0.16 | 4.49±6.40 | 23.16±34.86 | < 0.0001 |
| 3DU-Net | 0.66±0.20 | 0.68±0.20 | 8.81±8.38 | 49.70±93.40 | < 0.0001 |

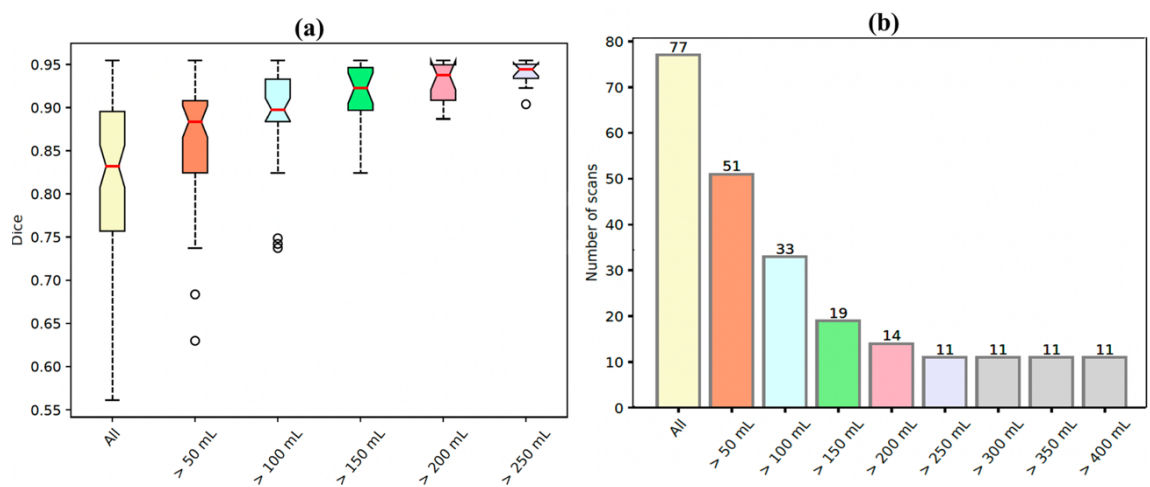

**Figure 3. Model performance with respect to effusion volume.** A. Per-class boxplots of Dice scores when including only PE with volume greater than the indicated thresholds. Central line, median. Boxes, interquartile range indicating 95% CI of the median. Whiskers represent 15 times the IQR. Blacked-outlined circles, outliers. B. The distribution of patients in each group.

In addition to computing the Dice coefficient across all test scans, we evaluated the TTAS' segmentation performance in relation to the pleural effusion (PE) volume (Fig. 4). The TTAS performed exceptionally well in segmenting large PE ($> 200$ mL), achieving a Dice coefficient of $0.93 \pm 0.01$ (Fig. 4a). For medium to large PE ($> 100$ mL), a Dice score of $0.89 \pm 0.06$ was attained, while for all PE cases, the Dice score reached $0.82 \pm 0.09$. Next, the qualitative segmentation results for three subjects using the three compared methods was compared (Fig. 5). In the visual representation, true positive, false negative, and false positive voxels are respectively depicted in blue, red, and green, in comparison to the corresponding ground truth segmentation. The TTAS method exhibited a significant reduction in false negatives and false positives when compared to the state-of-the-art methods. The conventional semi-supervised learning-based segmentation method Teacher-Student (TS) suffers from more false positive predictions, leading to an overestimation of the PE volume. In contrast, the proposed TTAS model generates an accurate segmentation mask and subsequently provides a more precise estimation of the PE volume.

**Table 2. Quantitative performance analysis with respect to the different number of training samples.** Means and standard deviations of the Dice coefficients on test scans when 50 and 25 annotated scans were used for training. The best results are indicated in bold font. Asterisks denote statistical significance ($p < 0.05$) in the difference between the proposed method and the competing method, as determined by a paired Student's t-test.

| Methods | 50 Annotated Training Scans | 25 Annotated Training Scans |
|---|---|---|
| TTAS | **0.76±0.06** | **0.72±0.10** |
| TS | 0.71±*0.12* | 0.66±*0.15* |
| nnU-Net | 0.62±*0.16* | 0.53±*0.18* |
| nnU-Net | 0.57±*0.21* | 0.51±*0.25* |

Finally, we investigated the impact of the number of training scans. Specifically, we explored two additional scenarios in which we used $50$ and $25$ randomly selected annotated scans from the AIDA-CTPA dataset to train the models. Subsequently, we conducted testing on the $77$ samples from the PleThora dataset. Additionally, as previously, we utilized the remaining unannotated 181 sample cases from AIDA-CTPA to train semi-supervised learning-based models, such as the proposed TTAS and TS.

The means and standard deviations of the Dice coefficients are showcased in the presented results (Table 2). Notably, the semi-supervised segmentation methods demonstrated strong generalization capabilities, benefiting from their capacity to learn from unlabelled data. In contrast, the supervised methods, specifically nnU-Net and 3DU-Net, struggled and exhibited poor generalization due to the limited size of the training set, which inadequately represents the distribution of annotated data. The proposed TTAS consistently achieved higher mean Dice coefficients in both experiments and remained stable despite the small training set size, outperforming the competing methods.

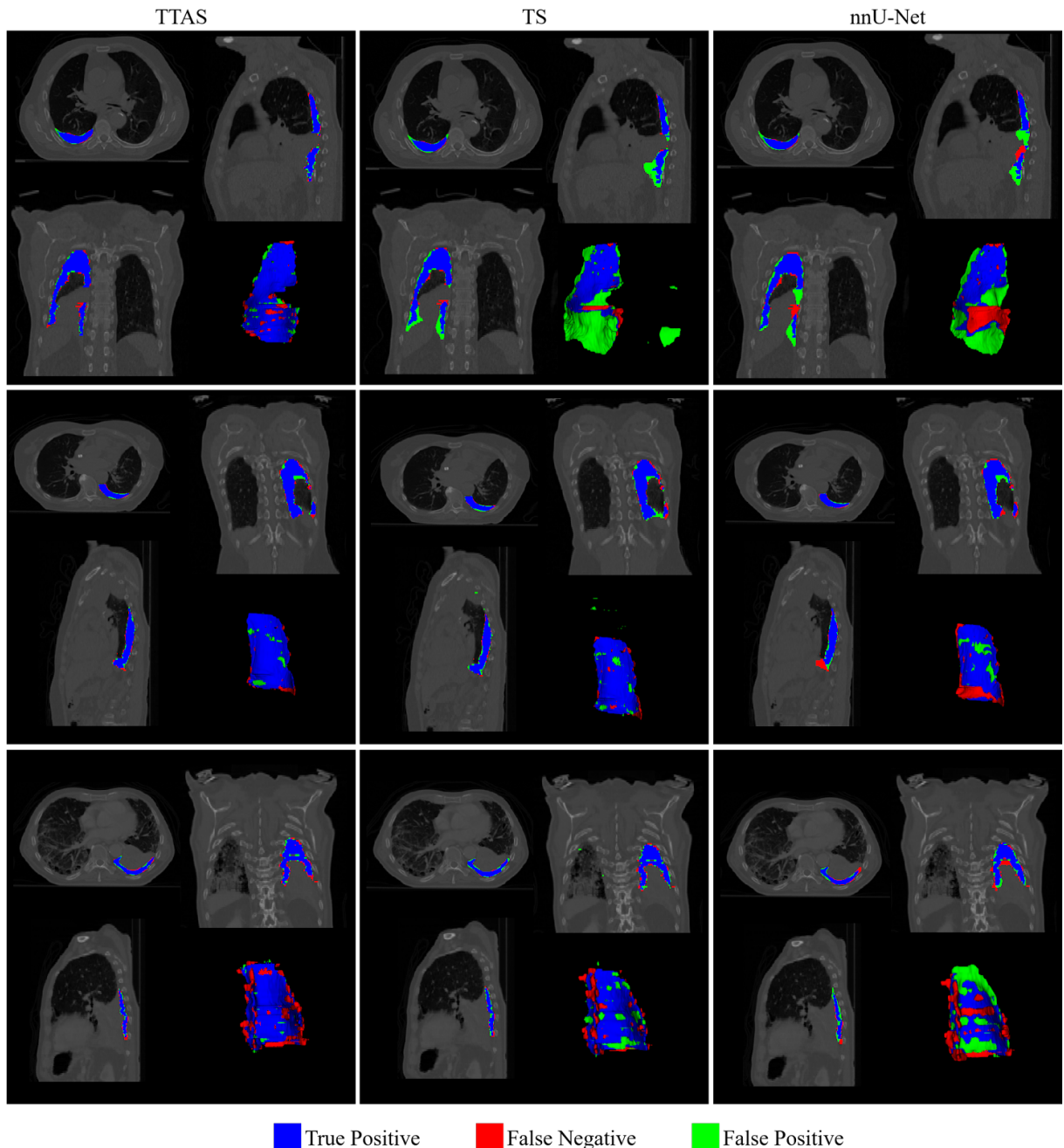

**Figure 4.** Qualitative segmentation results (axial, coronal, sagittal, and 3D views) for three sample patients. Accurately predicted voxels highlighted as true positives and incorrectly predicted voxels as either false positives or false negatives by comparing them with manual segmentations.

**Discussion**

Recent advancements in Artificial Intelligence (AI) have catalyzed the development of computer-aided detection and diagnosis systems in thoracic radiology, specifically enabling the accurate identification of PE from CT scans (6,8). The primary challenge in this field lies in the scarcity of a large and diverse annotated dataset that encompasses the broad spectrum of pathological pleural fluid presentations (10). The limited availability of labeled data poses a significant obstacle to training and evaluating automated methods, often resulting in suboptimal performance and restricted applicability in clinical practice (16). In such circumstances, semi-supervised learning-based segmentation methods prove invaluable. However, they may be susceptible to inaccurate mask predictions, often arising from learning unreliable pseudo-labels (17). The conventional semi-supervised learning-based segmentation method, Teacher-

Student (TS), is prone to more false positive predictions, leading to an overestimation of the PE volume (18). This paper introduces a novel semi-supervised deep learning framework, Teacher-Teaching Assistant-Student (TTAS), designed for the segmentation and quantification of PE from contrast-enhanced CT volumes. In contrast to existing methods, the proposed TTAS model demonstrates the capability to generate an accurate segmentation mask, subsequently providing a more precise estimation of the PE volume.

Our proposed TTAS method successfully addresses and resolves this issue by allowing the feature extractor to learn feature representations from both accurate labels and pseudo-labels, while the mask predictor exclusively learns from accurate labels. This approach results in more precise segmentation outcomes, achieving robust segmentation performance with a mean Dice coefficient of 0.82 (95% CI, 0.79 – 0.84). We rigorously evaluated the performance of our segmentation method on a publicly available clinical dataset PleThora (12), which comprises a wide range of PE of varying sizes, ranging from 22 mL to 446 mL, as well as cases without effusions. The TTAS algorithm delivers a robust volumetric measurement of pleural fluid with a significantly lower volumetric difference compared to manual reference segmentation, as indicated by the mean AbVD of 6.49 mL (95% CI, 4.80 – 8.20). Our results indicate that the TTAS model outperformed state-of-the-art models in key performance metrics, including the Dice coefficient, Precision, Average Surface Distance (ASD), and Absolute Volume Difference (AbVD). TTAS demonstrated strong generalization capabilities, benefiting from its capacity to learn from unlabeled data, which is particularly valuable when dealing with limited annotated data.

Manual segmentation of PE from CT volume is inherently challenging, even for experienced radiologists. As previously mentioned, a subset of the PleThora dataset contains manual segmentations provided by three radiologists for 16 cases. To gauge the level of agreement among these annotators, we computed the Intraclass Correlation Coefficient (ICC). For manual segmentation, the ICC was found to be 0.63, indicating moderate agreement among the radiologists. In contrast, when it comes to automated segmentation, we observed no significant inter-observer variability, with an ICC of 0.99. This underscores the consistency and reliability of the automated segmentation process. While it is evident from (Fig. 4) that TTAS excels in segmenting large PE cases (> 200 mL) a linear regression analysis revealed a moderate positive linear relationship (r : 0.59; p-value < 0.001) between the segmentation Dice score and the volume of PE.

The automatic detection and reliable segmentation of pleural effusions (PE) in chest CT scans enable routine, interaction-free utilization, 3-dimensional volumetric analysis, and swift quantification. Nevertheless, there are numerous aspects that warrant further exploration in the future, particularly involving a more extensive and diverse dataset. This dataset should encompass CT images collected from various healthcare facilities, utilizing scanners from different manufacturers and serving diverse patient demographics. We are also in the process of developing a web-based interface to provide convenient access to the volume quantification model, offering an interactive refinement feature for fine-tuning, if necessary, to obtain accurate Pleural Effusion (PE) volume estimates. Our ultimate goal is to seamlessly integrate this system with the Picture Archiving and Communication System (PACS) within the institute's hospital.

## References

1. Jeba, J., Cherian, R. M., Thangakunam, B., George, R. & Visalakshi, J. Prognostic factors of malignant pleural effusion among palliative care outpatients: A retrospective study. *Indian J. Palliat. Care* 24, 184 (2018).

2. Walker, S. P. *et al.* Nonmalignant pleural effusions: a prospective study of 356 consecutive unselected patients. *Chest* 151, 1099–1105 (2017).

3. McClain, L. *et al.* Admission chest radiographs predict illness severity for children hospitalized with pneumonia. *J. hospital medicine* 9, 559–564 (2014).

4. Hazlinger, M., Ctvrtlik, F., Langova, K., Herman, M. *et al.* Quantification of pleural effusion on ct by simple measurement. *Biomed Pap Med Fac Univ Palacky Olomouc Czech Repub* 158, 107–111 (2014).

5. Mergo, P. J. *et al.* New formula for quantification of pleural effusions from computed tomography. *J. thoracic imaging* 14, 122–125 (1999).

6. Sexauer, R. *et al.* Automated detection, segmentation, and classification of pleural effusion from computed tomography scans using machine learning. *Investig. Radiol.* 57, 552–559 (2022).

7. Yao, J., Han, W. & Summers, R. M. Computer aided evaluation of pleural effusion using chest ct images. In *2009 IEEE International Symposium on Biomedical Imaging: From Nano to Macro*, 241–244 (IEEE, 2009).

8. Isensee, F., Jaeger, P. F., Kohl, S. A., Petersen, J. & Maier-Hein, K. H. nnu-net: a self-configuring method for deep learning-based biomedical image segmentation. *Nat. methods* 18, 203–211 (2021).

9. Kahraman, F. T. T. D. e. a., A.T. Automated detection, segmentation, and measurement of major vessels and the trachea in ct pulmonary angiography. *Sci Rep 13, 18407* (2023).

10. Lee, J. H., Choi, C.-M., Park, N. & Park, H. J. Classification of pleural effusions using deep learning visual models:
contrastive-loss. *Sci. Reports* 12, 1–9 (2022).

11. Sjöblom, T., Sladoje, N., Kahraman, A. T., Toumpanakis, D. & Fröding, T. Computed tomography pulmonary angiography (ctpa) data. *N/A* DOI: 10.23698/aida/ctpa (2019).

12. Kiser, K. J. *et al.* Plethora: Pleural effusion and thoracic cavity segmentations in diseased lungs for benchmarking chest ct processing pipelines. *Med. physics* 47, 5941–5952 (2020).

13. Bakas, S. *et al.* Identifying the best machine learning algorithms for brain tumor segmentation, progression assessment, and overall survival prediction in the brats challenge. *arXiv preprint arXiv:1811.02629* (2018).

14. Qin, D. *et al.* Efficient medical image segmentation based on knowledge distillation. *IEEE Transactions on Med. Imaging* 40, 3820–3831 (2021).

15. Çiçek, Ö., Abdulkadir, A., Lienkamp, S. S., Brox, T. & Ronneberger, O. 3d u-net: learning dense volumetric segmentation from sparse annotation. In *Medical Image Computing and Computer-Assisted Intervention–MICCAI 2016: 19th International Conference, Athens, Greece, October 17-21, 2016, Proceedings, Part II 19*, 424–432 (Springer, 2016).

16. Candemir, S., Nguyen, X. V., Folio, L. R. & Prevedello, L. M. Training strategies for radiology deep learning models in data-limited scenarios. *Radiol. Artif. Intell.* 3, e210014 (2021).

17. Han, K. *et al.* An effective semi-supervised approach for liver ct image segmentation. *IEEE J. Biomed. Heal. Informatics* 26, 3999–4007 (2022).

18. Chaitanya, K., Erdil, E., Karani, N. & Konukoglu, E. Local contrastive loss with pseudo-label based self-training for semi-supervised medical image segmentation. *Med. Image Analysis* 87, 102792 (2023).

19. Lee, D.-H. Pseudo-label : The simple and efficient semi-supervised learning method for deep neural networks (2013).

20. Yang, L., Zhuo, W., Qi, L., Shi, Y. & Gao, Y. St++: Make self-training work better for semi-supervised semantic segmentation. In *Proceedings of the IEEE/CVF Conference on Computer Vision and Pattern Recognition (CVPR)*, 4268–4277 (2022).

21. Wang, Y. *et al.* Semi-supervised semantic segmentation using unreliable pseudo-labels. In *2022 IEEE/CVF Conference on Computer Vision and Pattern Recognition (CVPR)*, 4238–4247, DOI: 10.1109/CVPR52688.2022.00421 (2022).

22. Tarvainen, A. & Valpola, H. Mean teachers are better role models: Weight-averaged consistency targets improve semi-supervised deep learning results, DOI: 10.48550/ARXIV.1703.01780 (2017).